\documentclass[aps,prl,reprint]{revtex4-1}

\usepackage[ansinew]{inputenc}
\usepackage[english]{babel}
\usepackage{textcomp}
\usepackage{blindtext}
\usepackage{csquotes}
\usepackage{mathrsfs}
\usepackage{amsfonts}
\usepackage{mathtools}
\usepackage{amssymb}
\usepackage{amsmath}
\usepackage{amsthm}
\usepackage{amscd}
\usepackage{tensor}
\usepackage{stmaryrd}
\usepackage{rotating}
\usepackage[text={6.8in,8.5in},centering]{geometry}
\usepackage{tikz}
\usetikzlibrary{knots}
\usetikzlibrary{arrows, calc, decorations.markings, positioning}
\usepackage{tikz-cd}
\usepackage{relsize} 
\usepackage{hyphenat}
\usepackage{epigraph}
\usepackage{tikz-cd}

\tikzcdset{row sep/normal=0.6cm}
\tikzcdset{column sep/normal=0.2cm}

\tikzcdset{row sep/small=0.3cm}
\tikzcdset{column sep/large=0.6cm}

\tikzcdset{row sep/large=0.9cm}
\tikzcdset{column sep/small=0.1cm}

\usepackage{graphicx}
\graphicspath{ {./Images/} }

\usepackage{xcolor}
\usepackage[linktocpage]{hyperref}
\usepackage{makeidx}
\definecolor{nred}{rgb}{0.7,0,0}
\definecolor{nblue}{rgb}{0,0,0.6}
\definecolor{ngreen}{rgb}{0,0.6,0}

\hypersetup{
    colorlinks=true,
    linkcolor=nred,
    filecolor=nred,      
    urlcolor=ngreen,
    citecolor=nblue,
}

\begin{document}

\title{Beyond Binary:\\Hypermatrix Algebra and Irreducible Arity in Higher-Order Systems}
\author{Carlos Zapata-Carratal\'a}
\email{c.zapata.carratala@gmail.com}
\affiliation{SEMF, Wolfram Institute}
\author{Maximilian Schich}
\affiliation{CUDAN Lab, Tallinn University}
\author{Taliesin Beynon}
\affiliation{Mathematics Department, University of Cape Town}
\author{Xerxes D. Arsiwalla}
\affiliation{Pompeu Fabra University, Wolfram Research}

\maketitle

\sloppy

\section*{Abstract}

Theoretical and computational frameworks of modern science are dominated by binary structures. This binary bias, seen in the ubiquity of pair-wise networks and formal operations of two arguments in mathematical models, limits our capacity to faithfully capture irreducible polyadic interactions in higher-order systems. A paradigmatic example of a higher-order interaction is the Borromean link of three interlocking rings. In this paper we propose a mathematical framework via hypergraphs and hypermatrix algebras that allows to formalize such forms of higher-order bonding and connectivity in a parsimonious way. Our framework builds on and extends current techniques in higher-order networks -- still mostly rooted in binary structures such as adjacency matrices -- and incorporates recent developments in higher-ar ity structures to articulate the compositional behaviour of adjacency hypermatrices. Irreducible higher-order interactions turn out to be a widespread occurrence across natural sciences and socio-cultural knowledge representation. We demonstrate this by reviewing recent results in computer science, physics, chemistry, biology, ecology, social science, and cultural analysis through the conceptual lens of irreducible higher-order interactions. We further speculate that the general phenomenon of emergence in complex systems may be characterized by spatio-temporal discrepancies of interaction arity.

\textbf{Statement of Significance:} Binary structures are ubiquitous in modern theoretical and computational frameworks. This resonates with the lack of established research on higher-arity structures in the mathematical literature. Consequently, most mathematical descriptions across science stay limited to the use of fundamentally binary models. This is particularly problematic in the study of higher-order phenomena where the behaviour of a system cannot be broken down into pair-wise interactions. Studying higher-arity mathematical models, such as hypergraphs, hypermatrices and $n$-ary algebras, is thus of capital importance, promising substantial multidisciplinary advances in higher-order systems science.

\textbf{Keywords} - arity, higher-order system, complexity, emergence, non-reductionism, hypergraph, network science, category theory, n-ary algebra, multicomputation, nuclear physics, quantum entanglement, semiotics, mathematical cognition, protein interactions, genome topology, molecular computing, assembly theory, cultural analytics, interdisciplinary research

\tableofcontents

\maketitle

\section{Introduction}\label{intro}

Systems whose behaviour cannot be broken down into pairwise interactions are generally known as \textbf{higher-order systems}. The mathematical problem to characterize such systems has been in the limelight in the last few years due to the diminishing returns of decades-old network science based on binary graph models. The recent monograph `Higher-Order Systems' \cite{battiston2022higher} presents the state of the art of this fast-growing field that harnesses \textbf{hypergraphs}, generalizations of graphs where edges connect an arbitrary number of nodes, to capture higher-order phenomena. This coincides with a number of voices in multiple scientific fields, including theoretical physics \cite{kerner2008ternary,azcarraga2010nary,baez2011invitation,benini2021operads}, biology \cite{klamt2009hypergraphs,kempes2021multiple}, neuroscience \cite{zhou2006learning,yu2011higher}, ecology \cite{billick1994higher,mayfield2017higher,valverde2020coexistence}, complexity science \cite{baas2009new,courtney2016generalized,arsiwalla2016global,arsiwalla2018measuring,neuhauser2021consensus} and computer science \cite{zhou2006learning,moskovich2015tangle,wolfram2021multicomputation,arsiwalla2021homotopies,arsiwalla2021pregeometric,eliassirad2021higher}, all of which echo concerns about current paradigms due to their limited binary nature. The rise of higher-order systems science has been succinctly captured in the Quanta Magazine article `How Big Data Carried Graph Theory Into New Dimensions' \cite{quanta2021higher}.

A key aspect of higher-order models is the richer compositional structure induced by the multiple ways in which parts of a system can interact simultaneously \cite{baas2014higher,baas2015higher}. Accordingly, this should entail the use of \textbf{higher-arity mathematical operations}, i.e. those that fundamentally act on an arbitrary number of arguments. However, what we find in most of the current research on higher-order systems is a mere extension of the mathematical techniques of network theory, still rooted in binary structures, to the study of hypergraphs \cite{battiston2021physics} and simplicial complexes \cite{courtney2016generalized}. We shall take a parsimonious approach to compositionality via a mathematical framework that successfully captures the minimal complexity of higher-order systems, going on to argue that such a formalization of higher-order compositionality is yet to be developed.

Our proposal makes use of \textbf{hypermatrices} \cite{kerner2008ternary,gnang2020bhattacharya,gnang2021hypermatrix,zapata2022heaps} and operations involving an arbitrary number of arguments, typically known in the mathematical literature as \textbf{$n$-ary algebras} \cite{carlsson1980n,filippov1985n,markl2002operads}. Hypermatrices are simply indexed arrays of algebraically active elements, such as numbers, truth values, etc. We must warn the reader that the precise mathematical definitions of `array' and `tensor' are not interchangeable, despite these terms being used as synonyms in the machine learning literature \cite{bratseth2021tensor}. As much as we would like to repurpose an existing mathematical toolkit and simply adapt it to our needs, it turns out that, historically, higher-arity structures have been largely ignored by the mathematical community \cite{corry2003modern} -- almost to a surprising degree. Consequently, developing an appropriate formalism of higher-order compositionality will require a degree of mathematical inventiveness uncommon in areas outside of pure mathematical research. Nevertheless, our work aligns with a growing interest in higher-order ideas across multiple branches of mathematics, as seen in the development of higher category theory \cite{lurie2009higher,simpson2011homotopy,arsiwalla2021homotopies,arsiwalla2021pregeometric}, operads \cite{markl2002operads,leinster2004higher}, opetopes \cite{baez1998higher,cheng2004weak}, tensor networks \cite{biamonte2017tensor}, hypercompositional algebra \cite{davvaz2009generalization,massouros2021overview}, heap theory \cite{kolar2000heap,hollings2017wagner, zapata2022heaps} and hyperstructures \cite{baas2019mathematics,baas2019philosophy}.

The central notion that our framework aims to put forward as a measure of atomic complexity in a higher-order system is that of \textbf{irreducible arity}: the minimal size of a collection of elements, particles, vortices, agents or, generally, parts of a system, that determines the nature of their mutual interaction. This concept is introduced in detail below and it is shown in context across natural systems such as atomic nuclei, biomolecules, cognition and computation, ecosystems, societies, and cultural products, including representations of knowledge.

\begin{figure}
    \centering
    \includegraphics[scale=0.45]{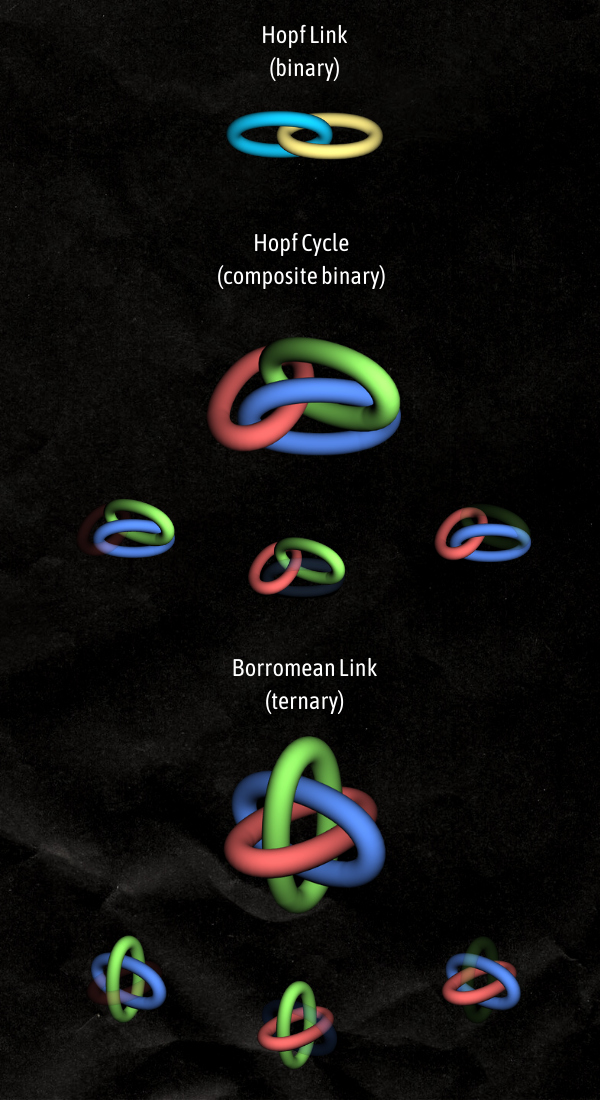}
    \caption{Illustration of some linkedness relations. Transparency is used to highlight the fact that the component loops of the Hopf cycle are pairwise linked while the loops in the Borromean configuration are pairwise unlinked.}
    \label{fig:LinkTable}
\end{figure}

\section{What is Arity?}\label{arity}

The word `arity' \cite{wiki2022arity} is a noun derived from the adjectives `binary', `ternary', `n-ary', etc., typically used to denote the number of elements in a relation or the number of arguments of an operation. Other terms such as `order', `degree', `adicity', `valency', `rank' or `dimension' are sometimes used to refer to similar concepts. We prefer the term `arity' given its emphasis on mathematical compositionality and the fact that it is seldom used outside of abstract algebra and relational structures, in contrast with the aforementioned terms that have well-established uses across mathematical sciences and beyond.

Arity may be generally understood as an elementary quantitative property of the interactions between parts of a system. In this sense, arity appears as a rudimentary measure of \emph{atomic complexity} or \emph{minimal emergence}. Let us clarify the notion of arity via an illustrative example. Consider a system of tangled loops of string and the linkedness relations defined on collections of loops by the condition that they stay bound together when each loop is mechanically pulled apart, that is, a physical analogue of mathematical links in knot theory. The smallest linkedness relations are binary since they involve two loops. For instance, the linkedness relation with the simplest topology is known as the Hopf link (Figure \ref{fig:LinkTable}, top). This configuration can be concatenated to form higher-arity relations. Given three loops, we can form the Hopf cycle (Figure \ref{fig:LinkTable}, centre). The condition that three loops form a Hopf cycle is a ternary linkedness relation that is decomposable into pairs of Hopf link relations. We thus say that the Hopf cycle is of arity 3 which is \textbf{reducible} into the lower arity 2. In contrast, we also find the Borromean link configuration (Figure \ref{fig:LinkTable}, bottom). The condition that three loops form a Borromean link is a ternary linkedness relation that cannot be decomposed into smaller linkedness relations of its constituents. We thus say that the Borromean link is of \textbf{irreducible} arity 3. More generally, loops forming Brunnian links give higher irreducible linkedness relations \cite{baas2014higher}.  The Borromean link is the paradigmatic example of the general phenomenon of \textbf{irreducible arity}: the minimal size of a collection of elements, particles, agents or, generally, parts of a system, that determines the nature of their mutual interaction. In the interest of brevity, unless otherwise specified, the term `arity' will mean `irreducible arity'. Thus, we shall refer to the Hopf cycle as a (composite) binary relation and to the Borromean link as a ternary relation.

\section{The Mathematics of Higher-Order Systems}\label{math}

\begin{figure}
    \centering
    \includegraphics[scale=0.45]{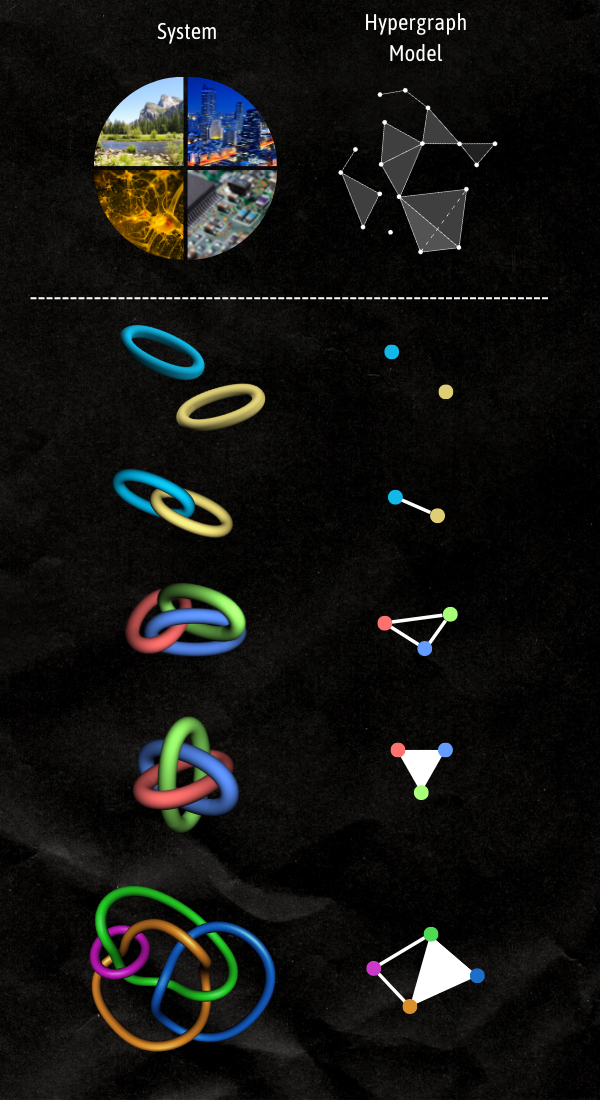}
    \caption{Tangles of loops, here representing general higher-order interactions among parts of a system, are associated with hypergraphs by identifying each loop with a distinct node and each topological link between them with an (undirected) hyperedge of order equal to the arity of the link.}
    \label{fig:SystemTable}
\end{figure}

The recent literature on higher-order systems \cite{zhou2006learning,klamt2009hypergraphs,courtney2016generalized,battiston2020networks,neuhauser2021consensus,battiston2021physics,arsiwalla2021homotopies,arsiwalla2021pregeometric,battiston2022higher} has introduced hypergraphs and simplicial complexes as generalizations of pairwise networks that can account for interactions of arbitrary arity. We shall focus only on \textbf{hypergraphs}, as simplicial complexes are particular cases of hypergraphs endowed with additional structure. Our goal in this section is to demonstrate the adequacy of hypergraphs to model higher-order systems with irreducible interactions and to show that an order-agnostic approach to adjacency properties of hypergraphs leads naturally to higher-arity algebras of hypermatrices.  Consider again the example of a system of tangled loops. Figure \ref{fig:SystemTable} shows how the topology of a tangle can be systematically captured via hypergraph data. Note that the presence of higher-arity links in the tangle demands higher-order hyperedges, a pairwise network model would be insufficient. We can see this explicitly in the different hypergraphs that result from the Hopf cycle and the Borromean link: the former gives a cycle of binary edges while the latter gives a single ternary hyperedge. One of the key advantages of hypergraphs is that they naturally encode pairwise and higher-order data simultaneously, as illustrated by the the bottom tangle in Figure \ref{fig:SystemTable}.

\begin{figure}
    \centering
    \includegraphics[scale=0.46]{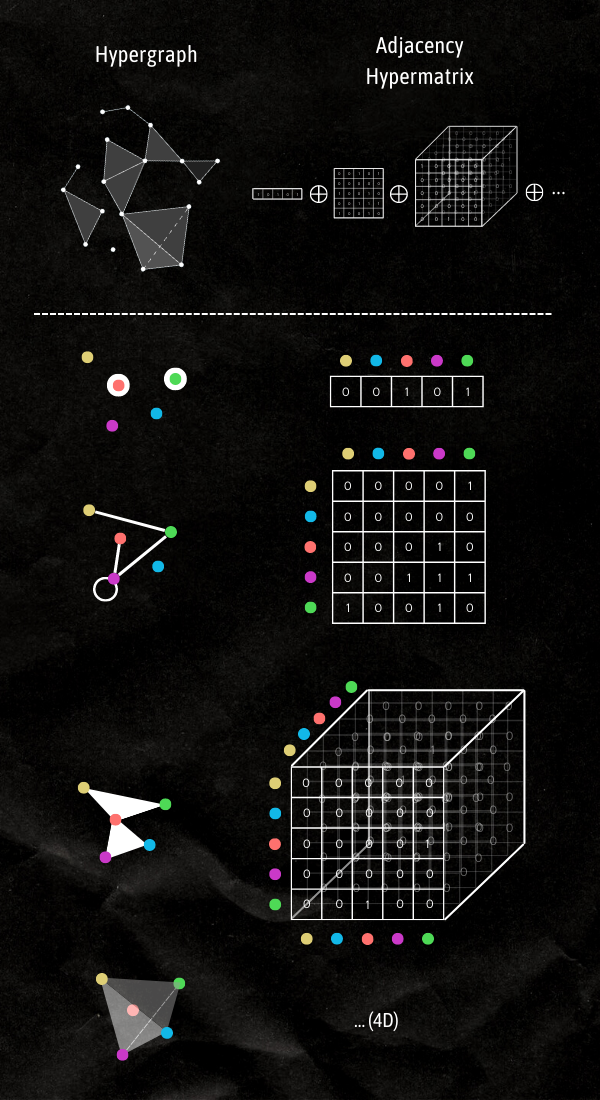}
    \caption{Visualization of Boolean-valued adjacency hypermatrices of orders $1$, $2$ and $3$. Higher-order hypermatrices cannot be represented by orthogonal arrays due to the dimensional limitations of 3D space, however they can be regarded as arrays of lower-order arrays.}
    \label{fig:MatrixTable}
\end{figure}

Let us focus on ordinary pairwise networks momentarily. The success of binary graphs as network models can be largely attributed to their capacity to encode \textbf{adjacency} data. At the lowest level, adjacency in binary graphs is simply the condition that nodes share a common edge. More generally, adjacency captures information about the connectivity structure (paths, distance, motifs...) between pairs of nodes. Adjacency data is made operational via the \textbf{adjacency matrix}: by labelling the nodes of a graph with some index variable $i\in [1,N]$ we define the entries of the adjacency matrix $A_{ij}$ according to whether nodes $i$ and $j$ are adjacent. Depending on the kind of graph data one wants to capture, the assignment of adjacency matrix entries can be defined in different ways: the simplest assignment is to take the binary Boolean algebra $(\{0,1\},\vee, \wedge)$ and set $A_{ij}=1$ when there exists at least one edge between nodes $i$ and $j$, and $A_{ij}=0$ otherwise (Figure \ref{fig:MatrixTable}, second row); we can also take the natural numbers $(\mathbb{N},+,\cdot)$ and set $A_{ij}$ to be the number of edges that exist between nodes $i$ and $j$; if the graph is weighted over continuous variables, to capture flows or reaction rates for instance, we take the real numbers $(\mathbb{R},+,\cdot)$ and set $A_{ij}$ to be the signed sum of all the weights between nodes $i$ and $j$. In all cases, the entries of the adjacency matrices are elements of a semiring, that is, an abstract set with additive and multiplicative operations $(S,+,\cdot)$ which can be understood as the minimal algebraic setting that enables basic arithmetic.

For general hypergraphs a similar construction is possible if we parse the set of hyperedges by order: for a fixed order $n$, the entry $A^{(n)}_{i_1i_2\cdots i_n}$ is defined according to the existence of a hyperedge between the nodes labelled by $i_1$, $i_2$, $\dots$ $i_n$; this results in a $n$-matrix $A^{(n)}$ that captures the adjacency information at order $n$. The \textbf{adjacency hypermatrix} of a hypergraph is given by the direct sum of all the fixed-order $n$-matrices:
\begin{equation*}
    A:=\bigoplus_{n=1}^\infty A^{(n)}.
\end{equation*}
This construction is illustrated in Figure \ref{fig:MatrixTable} for a small set of nodes and Boolean-valued adjacency hypermatrices of order up to $3$.
\begin{figure}
    \centering
    \includegraphics[scale=0.46]{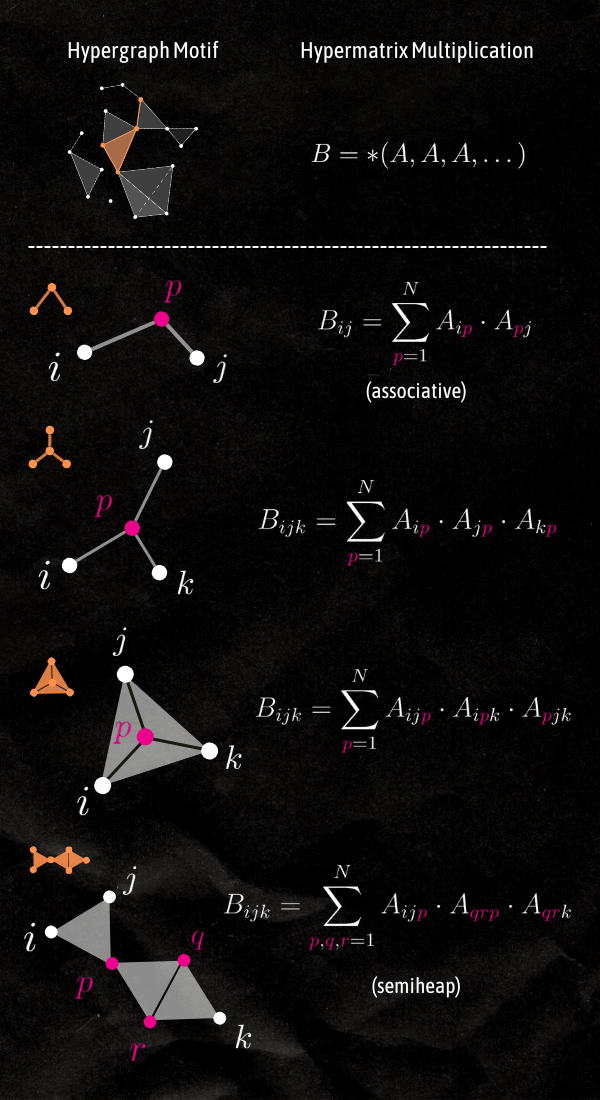}
    \caption{Several examples of hypermatrix multiplications defined from motif-adjacency in hypergraphs. External nodes (white) are fixed and the condition that there exists a subhypergraph matching the motif (orange) containing some internal nodes (red) is captured by the algebraic expression on the right in terms of the adjacency hypermatrix $A$.}
    \label{fig:MultiplicationTable}
\end{figure}

The power of the adjacency matrix -- which is a particular case of adjacency hypermatrix -- resides in its capacity to turn complicated graph topology questions into computationally efficient matrix algebra \cite{estrada2015network}. This is achieved by defining an operation on adjacency matrices from the condition of whether two nodes are connected via some intermediary node (Figure \ref{fig:MultiplicationTable}, first row). This recovers ordinary matrix multiplication, a binary operation that constitutes the prime example of typical compositional structures, such as categories \cite{lurie2009higher,simpson2011homotopy} and operads \cite{markl2002operads,leinster2004higher}, where objects are composed according to the associative rule:
\begin{equation*}
    (a(bc))=((ab)c).
\end{equation*}
Figure \ref{fig:MultiplicationTable} shows that this is a particular case of hypergraph \textbf{motif-adjacency}: given a motif (an isomorphism class of simple connected hypergraphs) the adjacency condition between a collection of nodes is defined by the existence of a matching subhypergraph sitting on the collection of nodes \cite{zapata2022invitation}. Ordinary matrix multiplication in linear algebra corresponds to the choice of a `V' motif (Figure \ref{fig:MultiplicationTable}, first row) while the choice of a `Y' motif (Figure \ref{fig:MultiplicationTable}, second row) leads to a multiplication that takes $2$-matrices and yields a $3$-matrix via a $3$-way index summation, an uncommon index operation in conventional linear algebra. Considering small order $3$ motifs we obtain higher-arity analogues of matrix multiplication. The cone motif (Figure \ref{fig:MultiplicationTable}, third row) results in a ternary operation, again involving a $3$-way index contraction, called the Bhattacharya-Mesner product \cite{mesner1990association,mesner1994ternary} for which no associativity-like properties are known. The fish motif (Figure \ref{fig:MultiplicationTable}, fourth row) results in a ternary operation satisfying the semiheap associativity property \cite{hollings2017wagner,zapata2022heaps}:
\begin{equation*}
    ((abc)de)=(a(dcb)e)=(ab(cde)).
\end{equation*}
We have thus shown that a general approach to higher-order adjacency properties of hypergraphs leads naturally to higher-arity matrix algebra.

Although higher-order matrices \cite{cayley1894collected} and relations \cite{peirce1870description,peirce1880algebra} have been known for more than 150 years, non-binary instances of such structures have received very little attention. In fact, higher-arity algebra \cite{dudek2007remarks,azcarraga2010nary,hollings2017wagner,rybolowicz2021topics,zapata2022heaps,rybolowicz2022biunit} and hypermatrix theory \cite{kerner1997cubic,kerner2008ternary,gnang2014combinatorial,gnang2020bhattacharya,gnang2021hypermatrix} are active fields of research that have only started to grow significantly in the last couple of decades. It is our belief that the correct modelling, and eventual understanding, of irreducible interactions in higher-order systems requires the development of a solid foundation of higher-arity hypermatrix algebra techniques which, given the state of the art, can only be attained via fundamental mathematical research.

\section{Irreducible Arity Across Science}\label{science}

Having introduced the mathematical machinery that articulates the abstract concept of irreducible arity, we now turn to the discussion of known instances of irreducible higher-order interactions in nature. There is a broad spectrum of complex systems where elementary cells of interaction involve more than a pair of agents: fundamental forces in multi-particle systems \cite{cohen1993fifty,dobnikar2002many}, processing nodes in computational frameworks \cite{andrews2000foundations,wolfram2021multicomputation}, primitive chemical reactions in metabolic networks \cite{wagner2001small,ravasz2002hierarchical}, inter-species relations in ecosystems \cite{schoenly1991trophic,billick1994higher,sole2012self,valverde2020coexistence}, protein interactions in living cells \cite{phizicky1995protein,nooren2003diversity,bertoni2017modeling}, etc. Irreducible arity is implicit in integrated information theory \cite{arsiwalla2013integrated,arsiwalla2016global,mediano2022integrated}, it appears as bonds in higher-order organization \cite{baas2013structure,baas2015higher}, it occurs across different coarse-graining scales \cite{flack2017coarse}, it gives rise to the notion of individual when persistent over time \cite{krakauer2020individuality}, and it is implicit in the `n-ary' building blocks constituting large collections of knowledge \cite{pellissier2016freebase}. Furthermore, the general phenomenon of emergence \cite{anderson1972more} may be hypothesized to be a manifestation of arity discrepancies across spatio-temporal scales. Beyond these general remarks, we present a few concrete examples of well-documented higher-order phenomena in systems where we can identify irreducible higher-arity interactions. This list serves as a testament to the transdisciplinary nature of the concept of irreducible arity.

\subsection{\textbf{Symbiosis and Ecosystems}}\label{eco}

For many decades, inter-species relations in ecosystems have been reported as paradigmatic examples of higher-order interactions \cite{abrams1983arguments,schoenly1991trophic,billick1994higher,wootton1994pieces,sole2012self}. In fact, the ecology community has been at the forefront of the development and application of higher-order techniques \cite{mayfield2017higher}, and has long warned about the shortcomings of conventional models that are limited to pairwise relations \cite{levine2017beyond,letten2019mechanistic}. More concretely, cases of irreducible multi-species interdependence have been recently characterized with hypergraph models \cite{valverde2020coexistence} and 3-way symbiotic relations have been directly observed \cite{marquez2007virus,roopin2011amonia}.

\subsection{\textbf{Efimov States and Borromean Nuclei}}\label{efimov}

The discovery of quantum three-particle bound states by Efimov in 1970 \cite{efimov1970energy} galvanized the field of few-body physics, a discipline that has seen continuous growth since \cite{hammer2010efimov,naidon2017efimov}. The Efimov effect, as it is know today, predicts states of three quantum particles that are bound in a trio while the pair configurations are unbound. This resulted in the search for experimental evidence of so-called Efimov states which led to the discovery of Borromean nuclei: a certain group of light nuclei characteristic for having excess neutrons that exhibit the precise scattering behaviour predicted by the Efimov effect \cite{cornelius1986efimov,fedorov1994efimov, bhasin2021few}.  
Borromean nuclei, as their technical name suggests, constitute direct evidence of irreducible ternary and higher-arity interactions in nuclear few-body systems. It is perhaps no coincidence that the first attempts to use higher-arity structures in physics were also in the context of nuclear interactions, as ternary Poisson brackets were considered by Nambu \cite{nambu1973generalized,nambu1974three} in an attempt to give a canonical quantization of the strong nuclear force. Borromean nuclei in interaction are a prototypical example of two classes of higher-order systems: in the many-particle limit they lead to higher-order statistical mechanics \cite{dobnikar2002many,dobnikar2004three,ampatzoglou2021rigorous}, and they are an instance of higher-order quantum entanglement \cite{you2020higher}.

\subsection{\textbf{Molecular Topology}}\label{chem}

In recent years there has been a significant development of synthesis techniques that allow the investigation and design of intricate nano-scale molecular structures \cite{stoddart2020dawning}. Remarkable examples of so-called nanotopology are catenanes \cite{gil2015catenanes}, rotaxanes \cite{schill2017catenanes} and molecular machines \cite{kay2015rise}. In parallel, research on biomolecules in living organisms has found a wealth of topologically non-trivial conformations in nucleic acids \cite{bon2008topological,huang2015shapes}, genomes \cite{chen1995topology}, and proteins \cite{taylor2003protein}.  The presence of higher-arity links in catenanes has been amply documented \cite{gil2015catenanes}, in particular, the existence of Borromean rings \cite{chichak2004molecular,meyer2010borromean}. Although direct evidence is lacking, higher-arity links are strongly suspected to occur in biological systems too. This topic is currently at the focal point of fields such as genome topology \cite{michieletto2015kinetoplast, main2021atomic} and protein entanglement \cite{niemyska2022alphaknot} in part due to the recent developments in atomic force microscopy \cite{krieg2019atomic} and the AlphaFold database \cite{david2022alphafold}. As per our discussion in previous sections, hypergraph and hypermatrix models would faithfully capture higher-order connectivity and are thus likely to prove instrumental in these lines of research.

\subsection{\textbf{Assembly and Biological Function}}\label{bio}

The general notion of assembly \cite{marshall2022formalising} typically presumes an additive framework in which components are attached one by one to form larger composite structures. However, there are systems where more than a pair of elements must be combined at a time to accomplish a desired structure. Examples of such systems are the so-called Borromean and Brunnian networks \cite{carlucci2003borromean,pan2014borromean}. The mathematical framework proposed by Assembly theory \cite{marshall2022formalising} is based on a binary graph model, therefore limited to only capture pairwise attachment, and requires the added data of an edge-valued function. A hypergraph model does away with these shortcomings by allowing assembly of arbitrary arity and by incorporating the edge-valued function into the hyperedge data.  
The phenomenon of higher-order assembly is of particular relevance for theoretical biology as it has already been captured in mathematical models of protein complex structure and function \cite{ortiz2020combinatorial}. As an example, the CD40L (CD154) protein trimer found in human cells \cite{song2015tnf} can only act as an ion channel when the precise trio of proteins is assembled, therefore the biological function of the complex is an irreducibly ternary relation between the proteins.

\subsection{\textbf{Cognition and Consciousness}}\label{cogn}

The particulars of human colour perception \cite{boynton1990human,king2005human} offer an intuitive source of ternary phenomena in the form of the experience of whiteness as emergent from the simultaneous combination of red, green and blue. Likewise, recent research on the ion channels involved in the nervous impulse for flavour and odour perception has shown that higher arity is present even at the structural molecular level: several molecules can engage with  multiple binding sites of a single ion channel simultaneously \cite{del2021structural}.

Considerations of holism and gestalt have also influenced recent  neuroscientific studies of consciousness, where a higher-order complexity, known as integrated information, has been postulated as a measure of consciousness \cite{tononi1998consciousness,tononi2004information,arsiwalla2013integrated}. In this view, conscious experience is quantified via the integrated information generated by the brain as a whole over and above the information generated by its parts    \cite{oizumi2014phenomenology,arsiwalla2016high,arsiwalla2016global}. This captures   processing complexity associated to simultaneous integration and differentiation of the brain's structural and dynamical motifs at all architectural scales   \cite{koch2016neural,arsiwalla2017morphospace,arsiwalla2018measuring}. The qualia of consciousness are represented as irreducible informational structures alluding to notions of higher-arity interactions \cite{tononi2016integrated}.

\subsection{\textbf{Generalized Computation}}\label{comp}

Modern computer science is undergoing a paradigm shift from single input-output sequences to distributed \cite{attiya2004distributed}, topological \cite{moskovich2015tangle}, multiway \cite{wolfram2002new,wolfram2021multicomputation,gorard2020zx,gorard2021zx,gorard2021fast,arsiwalla2021homotopies,arsiwalla2021pregeometric}, and collective \cite{kohler2022social} computational frameworks. At the same time, the development of diagrammatic quantum process algebra \cite{coecke2011interacting,coecke2018picturing} and the rise of topological quantum computing \cite{nayak2008non} have also contributed to this shift in perspective towards a generalization of the sequential nature of traditional computation. Perhaps the most explicit form of this trend is chemical and molecular computing \cite{sienko2003molecular,bhalla2014molecular}, where molecules and solutions of chemical compounds support well-defined discrete states and logic-like operations.

Conventional computation is abstracted into logic and relational algebra \cite{tarski1941calculus,fraisse2000theory}, which is commonly axiomatized via a combination of unary and binary operations. The generalized computational frameworks described above demand higher forms of logic that can capture irreducible concurrent interactions beyond pairs of computing elements. Higher-arity relations and their associated hypermatrix algebra \cite{zapata2022heaps} offer a natural way to encode such higher forms of logic.

\subsection{\textbf{Socio-Cultural Interaction and Knowledge Representation}}\label{collab}

Human societies are a rich source of instances of higher-order organization \cite{benson2016higher}. Collaborative activities, such as team assembly in scientific research \cite{newman2001structure,twyman2019team} or governance \cite{torfing2012governance} give rise to naturally irreducible groups of cultural production. At a smaller scale, relationships between individuals offer instances of low-arity affectional bonds \cite{ainsworth2006attachments} and small group dynamics \cite{homans2017human,balzarini2019demographic}. Non-human species have also been observed engaging in higher-order coordination, such as polyadic grooming in chimpanzees \cite{girard2020variable}. More general, higher-order interactions are well-documented in the literature on social networks \cite{scagliarini2021quantifying,battiston2020networks}. A further abundance of higher-order building blocks can be found in the documentation of cultural production and more general knowledge representation. Large knowledge graphs, for example, often routinely collect so-called `compound value types', higher-order `events', `n-ary' or `higher-arity relations' \cite{pellissier2016freebase,schich2010revealing,nickel2015review,hogan2021knowledge}. Both such explicit records and implicitly emerging higher-arity motifs constitute a mostly unclaimed, yet potentially highly fruitful area of multidisciplinary research, including social science, cultural analysis, and more general information science.

\section{Binary Bias and Arity Blindness}\label{binary}

Binary and sequential (composite binary) structures are so deeply ingrained in the fabric of modern science that noticing the prevalence of this particular form of arity can become a somewhat mind-bending exercise. Here are some examples of basic pieces of mathematical technology illustrating the ubiquity of binary structures throughout the literature:
\begin{itemize}
    \item   \textbf{Operations} ``$0+1$''. Ranging from elementary arithmetic to sophisticated notions in abstract algebra, operations almost exclusively take two arguments and most often they satisfy the associative rule, which makes them symbolically equivalent to sequences of characters.
    \item   \textbf{Relations} ``$x \sim y$''. Objects are often abstractly related in pairs, graphs being the minimal pairwise relational structure. Notably, equivalence relations and symbolic equality ``='', fundamental notions across mathematics and science, are binary relations.
    \item   \textbf{Processes} ``$A\to B$''. The time evolution of a system, the iteration of computational steps or the composition of transformations, usually encoded in functions and categories, rely on input-output or source-and-target paradigms.
    \item   \textbf{Language and Logic} ``\emph{cogito ergo sum}''. The spoken origins of written language make its sequential nature almost inevitable. When formalized into logic, unary or binary operations between sentences and propositions are sufficient to articulate constructions such as subject-verb-object, syllogisms and the standard reification into sequential triples source-link-target.
\end{itemize}
Although without further dedicated research we may only speculate the precise causes that led to this binary bias, we may offer some plausible explanations. In line with recent research on mathematical cognition \cite{nunez2017ancestral,nunez2017there} and cognitive anthropology \cite{overmann2021material}, one could surmise that the bilateral symmetry of human anatomy, the perception of handedness (chirality) or even the dual expression of sexual genotypes are contributing factors towards the preference for binary structures. A likely major influence in the prominent role of sequentiality can be found in language and education. Languages have been codified with strongly sequential systems that materially reflect primitive forms of verbal communication based on consecutive vocal cues \cite{dunbar2003social}. Those systems were incorporated into social traditions due to the development of education as a cultural phenomenon \cite{hogberg2015knowing} and they were further cemented by the invention of writing systems \cite{overmann2022early} and, eventually, printing systems \cite{gunaratne2001paper}.

In addition to the manifest preference for binary structures, the historical record shows that higher-arity structures have been largely neglected in their own right \cite{corry2003modern}. We suspect this to be the result of a lack of intuition for higher-order phenomena \cite{burr2010subitizing,clements2019subitizing} which, without the necessary embodied support and technology, can impede the identification of the correct arity of a structure under study, both in formal and natural sciences. We refer to this cognitive obstruction as \textbf{arity blindness}. In writing this paper it has been our aim to attempt to remedy our arity blindness by giving a transparent motivation of the mathematical framework that parsimoniously encodes arity (hypergraphs and hypermatrices) and offering an eclectic list of examples of natural systems that display higher-order behaviour and irreducible higher-arity interactions.

\section{Multidisciplinary Higher-Arity Science}\label{towards}

Despite the increasing popularity of higher-order models across scientific disciplines, most research is often limited to theories that merely extend conventional binary ideas. Indeed, network science, firmly based in the notion of binary graphs, is itself a relatively recent paradigm which is still permeating to some branches of natural sciences and the humanities. We believe that higher-arity concepts and techniques, such as our approach to higher-order hypergraph adjacency via hypermatrix algebra, should be fully embraced.

The many natural phenomena discussed in this paper offer ample evidence for the existence of a rich universe of genuinely higher-order behaviour awaiting to be explored.  The framework we propose sits at the intersection of two major trends in modern complexity science: on the one hand, its mathematical novelty and wide range of natural sciences to which it can be applied makes it a necessarily multidisciplinary effort, and, on the other, its parsimonious treatment of irreducible interactions positions it as an explicitly non-reductionist approach to higher-order systems modelling.  Binary bias and arity blindness create a sort of \emph{unthought frontier of science} that conceals potential discoveries and is a likely obstruction to many future breakthrough insights. A successful implementation of higher-arity methodologies in the spirit of our approach to irreducible higher-order interactions will have to contend with these obstacles. Notwithstanding, we believe this to be a unique opportunity that offers immense potential for mathematical creativity and scientific discovery, indeed in line with Leibniz's `similitudinis, ordinem, et relationibus expressionibus in universum'.

\bibliographystyle{apsrev4-1} % Tell bibtex which bibliography style to use
\bibliography{references.bib} % Tell bibtex which .bib file to use (this one is some example file in TexLive's file tree)

\end{document}